# Introduction to Physically Unclonable Fuctions: Properties and Applications

M. Garcia-Bosque, G. Díez-Señorans, C. Sánchez-Azqueta, S. Celma
Group of Electronic Design
Universidad de Zaragoza
Zaragoza, Spain
{mgbosque, gds, csanaz, scelma}@unizar.es

*Abstract*—During the last years, Physically Unclonable Functions (PUFs) have become a very important research area in the field of hardware security due to their capability of generating volatile secret keys as well as providing a low-cost authentication. In this paper, an introduction to Physically Unclonable Functions is given, including their definition, properties and applications. Finally, as an example of how to design a PUF, the general structure of a ring oscillator PUF is presented.

*Keywords*—PUF, physically unclonable functions, hardware security, authentication, secure key generation, random number generation

## I. INTRODUCTION

The necessity of transmitting information in a secure way has existed for a long time; however, this necessity has increased greatly in the last years, due to the exponential rise in digital communications. Nowadays, huge amounts of confidential and sensitive information are being transmitted at high speed and guarantying confidentiality for all this data constitutes a big challenge.

Traditionally, in the past, all the cryptographic primitives used to be analyzed only from a mathematical point of view, assuming that they behave as black boxes where possible attackers could only see the input and output, but not the internal operations. In the case of keyed primitives, it was usually assumed that it was easy to generate secure keys (secure key generation) and that they could be safely stored without being revealed (secure key storage). These assumptions, however, have proven to be very difficult to achieve from a practical point of view, requiring physical security measures. As a consequence, many security systems have been successfully attacked at a physical level. Regarding key generation, some systems have proven to be vulnerable [1], [2]. Regarding key storage, in order to be usable in an algorithm, keys are usually stored in non-volatile memories. Preventing well-equipped attackers to access to the content of these memories, have been proven to be very challenging [3], [4]. In this context, Physically Unclonable Functions (PUFs) constitute an excellent secure alternative to provide secure key generation and storage.

The idea behind a PUF is to use uncontrollable physical variations introduced during the manufacturing process to create physical entities with a unique and uncontrollable behavior. PUFs present a challenge-response functionality that is used to evaluate them. When a stimulus (challenge) is applied to a PUF instance, it behaves in an unpredictable but repeatable way. If the same challenge is applied to a different PUF instance, the response is different even if it was manufactured using the exact same process (Fig. 1). Due to this uniqueness, PUFs can be used for authentication purposes. On the other hand, using fuzzy extractors, they can also be used for secure generation and storage.

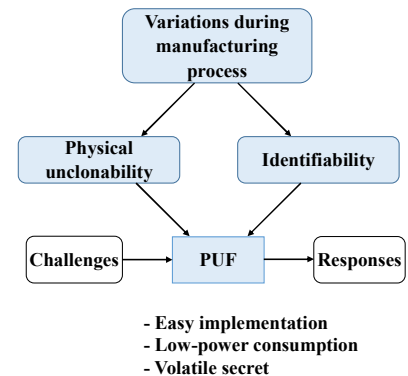

Fig. 1. General scheme of a PUF with its properties

Although several systems that exploited physical properties for authentication purposes had been proposed before, the first definitions of the PUF concept were proposed in 2001 and 2002 by Pappu [5], who introduced the term Physical One-Way Function (POWF), and Gassend et. al. [6] who used the term Physical Random Functions. However, despite being a relatively novel topic it has gained a lot of interest due to their applications in the field of hardware security. In this paper, we introduce the concept of PUF, explaining its main properties and applications for hardware security purposes. Furthermore, we present the scheme of the ring oscillator PUF (RO-PUF) to provide an example of how manufacturing process variations (which in this case produce small variations in the oscillation frequency) can be used to generate unique and measurable responses.

This paper is organized as follows: Section II presents the definition of PUF as well as its main properties; Section III explains how it is possible to use PUFs for authentication, key generation and storage applications; Section IV explains the

This paper has been supported by MINECO-FEDER (TEC2017-85867-R) and DGA fellowship to G. Díez-Señorans

general structure of RO-PUF; finally, conclusions are presented in Section V.

## II. DEFINITION AND PROPERTIES OF PUFs

### A. Definition

Along history, several definitions of PUFs have been given, attributing them certain properties. However, not all the proposed PUFs have met all these properties. In this work, we adopt the general definition proposed in [7]. This definition covers all the proposed PUFs and, at the same time, differentiates them from other constructions such as True Random Number Generators (TRNGs), RFID broadcasts or public-key signatures:

*Definition 1:* A Physically Unclonable Function (PUF) is a class of physical entities with a challenge-response functionality that exhibits identifiability and physical unclonability.

In the following subsections, the properties of identifiability and physical unclonability will be briefly explained along with other properties that, can be useful for some applications.

### B. Reproducibility

A PUF is said to be reproducible if any PUF instance always presents the same or very similar response when the same challenge is applied. The similarity between two responses for the same challenge and same PUF instance is called intra-distance and is usually measured using the Hamming Distance (HD). If the response of a PUF consists of an $m$-bit output word, $x = (x_1, x_2, \ldots x_m)$, the Hamming Distance between two responses $x, x'$ is defined as:

$$HD = \sum_{i=1}^{m} x_i \oplus x'_i \qquad (1)$$

The Hamming Distance is often expressed as a percentage.

Usually, the intra-distance depends on ambient conditions such as temperature or supply voltage. In these cases, it is necessary to measure the intra-distance response for different ambient conditions.

### C. Uniqueness

A PUF presents the uniqueness property if, when the same challenge is applied for different PUF instances, their responses are very different (i.e., their inter-distance is large). In a similar way as in the reproducibility case, inter-distances are often calculated using the Hamming Distance. However, uniqueness is usually evaluated under nominal operating conditions and is not evaluated under varying conditions.

### D. Identifiability

A PUF presents the identifiability property if PUF instances can be identified by measuring their responses. To meet these requirements, the PUF must meet the properties of reproducibility and uniqueness. It must be noticed that, to be identifiable, a PUF does not need to present perfect reproducibility or perfect uniqueness. In practice, it is almost impossible to design PUFs with an average intra-distance equal to 0%. Instead, intra-distances follow a probability distribution around a low value. In a similar way, average inter-distances typically follow a probability distribution around a value that is slightly different from 50%. To be identifiable, the necessary condition is that the intra-distances must be smaller than the inter-distances with a high probability.

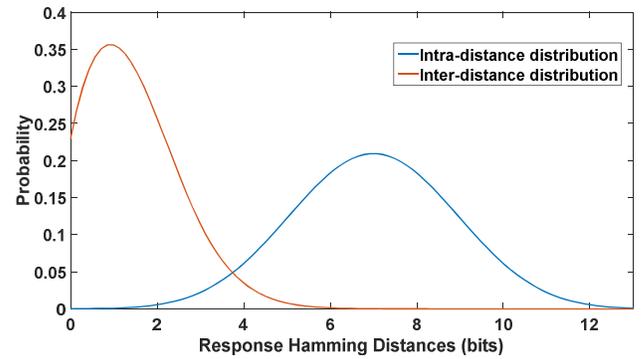

(a)

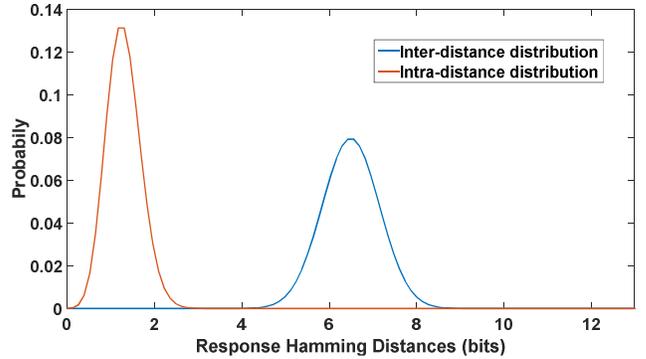

(b)

Fig. 2. Response Hamming Distance distributions for PUF with (a) low identifiability and (b) high identifiability

As an example, Fig. 2a shows a possible probability distribution of the Hamming Distances in a PUF. As it can be seen, if the measured Hamming Distance between two responses is 3 or 4 bits, it can be difficult to identify whether the responses correspond to the same PUF instance (in which case we would be measuring the intra-distance) or to a different PUF instance (in which case, we would be measuring the inter-distance). In this example, the PUF presents a low identifiability. On the other hand, if the probability distribution of the intra-distances and inter-distances almost do not overlap (Fig 2b), the identifiability is high.

### E. Physical unclonability

A PUF presents physical unclonability if an attacker or even the authorized manufacturer cannot influence the creation procedure of the PUF to create two instances with a similar behavior. More precisely, if it is not possible to create two different PUF instances whose inter-distance is smaller than the average inter-distance with a high probability. This can be expressed informally as: *it is easy to create a random PUF instance but hard to create a specific one* [7].

This property distinguishes PUFs from other constructions such as block ciphers or public-key signatures and, along with the identifiability property is used to defined the concept of PUF. The importance of this property relies on the fact that it is not necessary to trust the manufacturer since the construction process cannot be influenced to prevent each instance from being unique.

*F. One-wayness*

A PUF has one-wayness if it is difficult to invert. More precisely, given a PUF instance and a response, it is difficult to find a challenge that produces that response. In the classic definition of one-way functions, this property implies that the relationship between the inputs and outputs is complex so it is difficult to find an algorithm that is able to invert the function. However, in the case of PUFs, the one-wayness property also implies that the number of challenge-response pairs is big since, otherwise, an attacker could exhaust all the challenge-response pairs and create a lookup table that could be used to invert any response.

Although the first PUF proposed by Pappu [5] presented this property and he called his structure Physical One-Way Function, most of the PUFs proposed nowadays do not present this property. Furthermore, this property is not needed for most PUF security applications [8].

*G. Mathematical and True Unclonability*

A PUF presents Mathematical Unclonability if an attacker cannot predict the response of a random challenge, after having access to a big amount of challenge-response pairs. This property requires that the challenge set is large (typically escalates exponentially with some parameter such as the implementation area). Furthermore, it requires that an attacker cannot create an algorithm able to model the PUF from a set of challenge-response pairs. This property is difficult to achieve since most of the PUFs proposed so far have been proven to be vulnerable to machine learning attacks that are capable of modeling their behavior [9], [10].

The combination of mathematical and physical unclonability makes a PUF truly unclonable.

*H. Tamper Evidence*

Finally, a PUF presents tamper evidence when it is difficult to physically make small modifications in the behavior of a PUF instance. In the ideal case, any attempt to modify the behavior of a PUF instance results in a completely different behavior undistinguishable from a new unique PUF instance.

## III. APPLICATIONS

*A. Identification and Authentication*

Although some authors use identification and authentication as synonyms [11], other works make a distinction between both terms that we will use in this manuscript [7]. According to this approach, identification consists of stating the identity of an entity, without giving any proof. On the other hand, authentication is a stronger property that requires a proof of the identity of the entity. In other words, when a verifier needs to verify the identity of an entity, this entity must provide some proof of its identity (i.e., some secret that only that entity could know).

Using PUFs for identification purposes is not straightforward since, as explained before, PUF responses are not random uniformly distributed (average inter-distances are not 50%) and they are not reproducible when measured several times (intra-distances are not 0%). This is called a fuzzy behavior [12].

In a typical identification application, an identification threshold is used. This way, if the distances between two measured responses is below that threshold, both responses are considered to come from the same PUF instance. Otherwise, they are considered to come from different PUF instances. However, due to the probabilistic distribution of the intra-distances and inter-distances, it is possible that an identification error is made, specially when there is a large overlap between the intra-distance and inter-distances curves like in Fig. 2a.

There are two different kind of errors. If two responses come from the same PUF instance but are identified as coming from different PUF instances the error is called false rejection. The probability that it happens is called False Rejection Rate (FRR). If two responses come from different PUF instances but are identified as coming from the same PUF instance, the error is called false acceptance. The probability that this happens is called False Acceptance Rate (FAR). Both rates depend on the chosen threshold: if a high threshold is used, the FRR is low but the FAR is high whereas a decrease of the threshold results in an increase of the FRR and a decrease of the FAR. Often, the threshold is chosen in a way that both FRR and FAR are equal. However, some applications can require a high FRR or a high FAR; in those cases, the threshold is chosen accordingly.

To build a PUF-based authentication scheme, the following scheme is proposed in [13], [14]:

1. First, a verifier records the identity of every entity and collects several random challenge-response pairs. Then the verifier stores in a database the challenge response pairs corresponding to each entity's ID in a database.

2. When an entity requires to be authenticated, the verifier picks one of the challenge-response pairs corresponding to the entity's ID that have been previously stored in the database. Then, the verifier asks the entity for the response to that challenge and checks whether the response given by the entity is equal (or very similar) to the response stored in its database.

The main drawback of this scheme is that, if the PUF can be mathematically cloned, the verifier cannot distinguish the real entity from an attacker with a mathematical clone of the PUF. Therefore, this scheme only works for PUFs that are mathematically unclonable (true unclonable PUFs). For PUFs that are not mathematically unclonable, there are some alternative schemes such as the one proposed in [15]. A common approach of these schemes consist of applying a hash function to the responses and comparing these hashes instead of the responses. Since the responses present some fuzziness, it is necessary to apply some kind of error-correction technique so that the hash function is always applied to the exact same response within the same PUF instance.

*B. Key Generation and Storage*

Due to their unpredictable behavior, PUFs can be a good alternative to generate volatile keys (i.e., only exist when the device is running) [16]. In addition to authentication, key generation and storage is the other important application of PUFs. In order to generate reproducible keys, it is not possible to use the PUF responses directly because, as explained before, they are not perfectly reproducible (i.e., the response to the same challenge for the same PUF instance is not always exactly the same). On the other hand, output responses are not always perfectly random and, even if they are random, sometimes the key requires some specific constraints that the response does not satisfy. To solve these two issues, responses typically go through the following process.

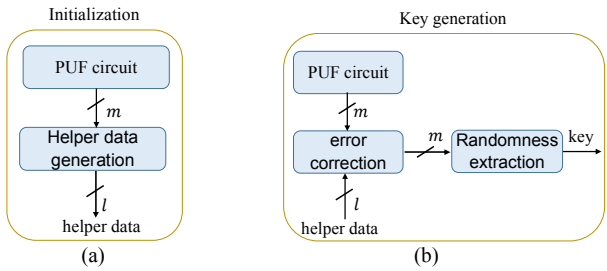

Fig. 3. Scheme of the process to generate secure cryptographic keys. (a) During an initialization phase, a helper data is produced so that the PUF always produce the exact same response. (b) During the key generation phase, the responses go through an error correction and randomness extraction function.

First, they pass through an error-correction function that guarantees that the same PUF instances always reproduce the exact same responses. To achieve this, a helper data that has been previously generated during an initialization phase is used.

Second, the corrected responses pass through a function that improves the randomness of the inputs and outputs secure cryptographic keys. The whole process is summarized in Fig. 3.

## IV. EXAMPLE: RO-PUF

A ring oscillator consists of an odd number of inverters connected in a loop. Its output oscillates at a frequency that, ideally, only depends on the number of inverters. In practice, however, due to random variations introduced during the manufacturing process, the oscillation frequency of each oscillator is not exactly the same. Typically, a RO-PUF compares the frequencies of pairs of identical oscillators to produce the output. A common scheme is shown in [16]. As it can be seen, the PUF contains an array of identical ring oscillators, a couple of multiplexers used to select the oscillators and a couple of frequency counters to measure the frequency of each oscillator. In this case, the challenge is the selection of the ring oscillators to compare and the response is the result of the comparisons.

Although there is a total of $\binom{n}{2}$ possible comparisons, not all of them produce independent outputs (if three oscillators have frequencies $f_1, f_2, f_3$ such us $f_1 > f_2$ and $f_2 > f_3$ it is straightforward than $f_1 > f_3$). To produce independent bits, there are several proposals such as using each oscillator once (which produces $n/2$ output bits) or comparing neighboring oscillators (which produces $n-1$ output bits) [17]. To further improve the reproducibility and uniqueness, some works divide the array in groups of $k$ oscillators and, within each group, only the pair of oscillators with the largest difference in frequency is compared (which produces $n/k$ output bits) [16].

## V. CONCLUSIONS

This paper has defined the concept of PUF and presented its main properties as well as other properties that can be useful for several applications. In addition, we have discussed how to exploit these properties for authentication and key generation or storage applications. Finally, as an example, the scheme of an RO-PUF has been presented to illustrate how the manufacturing process variations of a well-known structure (a ring oscillator) can be used to construct a PUF.